\documentclass[twocolumn]{aastex701}
\usepackage{placeins}

\usepackage{natbib}
\usepackage{amsmath}
\usepackage{lineno}
\usepackage{graphicx}
\usepackage{multirow}
\usepackage{float}

\newcommand{\masyr}{mas\,yr$^{-1}$}
\newcommand{\pmra}{$\mu_{\alpha}$}
\newcommand{\pmdec}{$\mu_{\delta}$}
\newcommand{\kms}{km\,s$^{-1}$}
\newcommand{\Msun}{M$_{\odot}$}

\shorttitle{Search for planets around Altair}
\shortauthors{Meshkat et al.}

\begin{document}

\title{Searching for Planets around Altair with JWST/NIRcam}

\author[orcid=0000-0001-6126-2467,gname='Tiffany',sname='Meshkat']{Tiffany Meshkat}
\affiliation{NASA Exoplanet Science Institute, IPAC, Caltech, MC 100-22, 1200 E. California Blvd., Pasadena, CA 91125, USA} 
\affiliation{IPAC, Caltech, MC 100-22, 1200 E. California Boulevard, Pasadena, CA 91125, USA}
\email[show]{meshkat@ipac.caltech.edu}  

\author[orcid=0000-0002-5627-5471,gname='Charles',sname='Beichman']{Charles Beichman}
\affiliation{NASA Exoplanet Science Institute, IPAC, Caltech, MC 100-22, 1200 E. California Blvd., Pasadena, CA 91125, USA} 
\affiliation{Jet Propulsion Laboratory, California Institute of Technology, Pasadena, CA 91109, USA}
\email{chas@ipac.caltech.edu}

\author[orcid=0000-0002-3414-784X,gname='Jorge',sname='Llop-Sayson']{Jorge Llop-Sayson}
\affiliation{Jet Propulsion Laboratory, California Institute of Technology, Pasadena, CA 91109, USA}
 \email{jorge.llop.sayson@jpl.nasa.gov}
 
\author[orcid=0000-0001-5966-837X,gname='Geoffrey',sname='Bryden']{Geoffrey Bryden}
\affiliation{Jet Propulsion Laboratory, California Institute of Technology, Pasadena, CA 91109, USA}
\email{geoffrey.bryden@jpl.nasa.gov}
 
\author[orcid=0000-0002-7162-8036,gname='Alexandra',sname='Greenbaum']{Alexandra Z. Greenbaum}
\affiliation{IPAC, Caltech, MC 100-22, 1200 E. California Boulevard, Pasadena, CA 91125, USA}
\email{azg@ipac.caltech.edu}

\author[orcid=0000-0003-2008-1488,gname='Eric',sname='Mamajek']{Eric E. Mamajek}
\affiliation{Jet Propulsion Laboratory, California Institute of Technology, Pasadena, CA 91109, USA}
\email{eric.mamajek@jpl.nasa.gov}

\author[orcid=0000-0001-7591-2731,gname='Marie',sname='Ygouf']{Marie Ygouf}
\affiliation{Jet Propulsion Laboratory, California Institute of Technology, Pasadena, CA 91109, USA}
\email{marie.ygouf@jpl.nasa.gov}

\author[orcid=0000-0001-8627-0404,gname='Julien',sname='Girard']{Julien Girard}
\affiliation{Space Telescope Science Institute, 3700 San Martin Dr, Baltimore, MD 21218, USA}
\email{jgirard@stsci.edu}

\author[orcid=0000-0002-0834-6140,gname='Jarron',sname='Leisenring']{Jarron Leisenring}
\affiliation{Steward Observatory, University of Arizona, Tucson, AZ 85721, USA}
\email{jarronl@arizona.edu}

\author[orcid=0000-0002-1838-4757,gname='Aniket',sname='Sanghi']{Aniket Sanghi}
\altaffiliation{NSF Graduate Fellow}
\affiliation{Cahill Center for Astronomy and Astrophysics, California Institute of Technology, 1200 E. California Boulevard, MC 249-17, Pasadena, CA 91125, USA}
\email{asanghi@caltech.edu}

\author[orcid=0000-0002-3191-8151,gname='Marshall',sname='Perrin']{Marshall Perrin}
\affiliation{Space Telescope Science Institute, 3700 San Martin Dr, Baltimore, MD 21218, USA}
\email{mperrin@stsci.edu}

\author[orcid=0000-0003-1227-3084,gname='Michael',sname='Meyer']{Michael Meyer}
\affiliation{Department of Astronomy, University of Michigan, 1085 S. University Ave., Ann Arbor, MI 48109, USA}
\email{mrmeyer@umich.edu}

\author[orcid=0000-0003-3818-408X,gname='Laurent',sname='Pueyo']{Laurent Pueyo}
\affiliation{Space Telescope Science Institute, 3700 San Martin Dr, Baltimore, MD 21218, USA}
\email{pueyo@stsci.edu}

\author[orcid=0000-0001-6396-8439,gname='William',sname='Balmer']{William Balmer}
\affiliation{Space Telescope Science Institute, 3700 San Martin Dr, Baltimore, MD 21218, USA}
\email{wbalmer@stsci.edu}

\author[0000-0001-8612-3236]{Andr\'{a}s G\'{a}sp\'{a}r}
\affiliation{Steward Observatory, University of Arizona, Tucson, AZ, 85721}
\email{agaspar@arizona.edu}

\author[0000-0003-0786-2140]{Klaus Hodapp}
\affiliation{University of Hawaii, Hilo, HI, 96720}
\email{hodapp@hawaii.edu}

\begin{abstract}
Altair ($\alpha$ Aql) is one of the nearest stars to the Sun at 5.13 pc and its age has recently been revised to 88$\pm$10 Myr, making it one of the youngest stars in the Solar neighborhood. This makes Altair a compelling target for high-contrast direct imaging searches for giant planets. We present two epochs of simultaneous F444W and F200W MASK430R JWST/NIRCam coronagraphic imaging observations. After post-processing, these data achieve F444W contrast levels of 2 x 10$^{-6}$ at 1'' (5.1 au), 1.5 x 10$^{-7}$ at 2'' (10.3 au), and a few x 10$^{-8}$ beyond 4''(20 au). Using the Sonora Flame Skimmer models, this corresponds to sub-Jupiter mass sensitivity beyond 5 au and sub-Saturn mass sensitivity beyond 10 au. We identified several candidate point sources in the field in the first epoch data. Second epoch observations confirmed that none of the candidates were co-moving with Altair and their colors are consistent with background stars. Altair has no known companions, no detected debris disk, and is not associated with any stellar association. Despite this, its youth and proximity ensure it will remain a high-priority target for future high-contrast imaging observations. We additionally discuss Altair as a potential reference star for future observations, presenting measurements of background sources in the field and showing that Altair will remain at least a few arcsec from background stars over the next 10–20 years, making it a reliable reference target for upcoming imaging campaigns including the Roman Space Telescope Coronagraph Instrument.
\end{abstract}


\section{Introduction} \label{Sec:intro}

The bright ($V=0.77$) A7V-type star Altair ($\alpha$ Aql, HD 187642, HIP 97649) is the second nearest ($\varpi = 194.95\pm0.57$\, mas, $d = 5.130\pm0.015$\, pc)
A-type star after Sirius \citep{Mermilliod97,Gray03,vanLeeuwen07}.
Thus far, Altair has no known stellar or substellar companions. 
Its rapid rotation (v sin i =221 km s$^{-1}$, \citealt{Erspamer2003}) and extreme brightness makes radial velocity and transit planet searches impractical, and no companion has been detected in interferometric observations \citep{Monnier07, Bouchaud20}. Coronagraphic imaging remains a promising method to discover companions at wider separations. Despite having an entry in the Washington Double Star Catalog (WDS J19508+0852A) and having a half dozen associated companions identified B through G with separations ranging from 27\arcsec\, to 196\arcsec, all appear to be faint, low proper motion background stars inconsistent with Altair's proximity or motion (see  \autoref{sec:siblings}). 

Altair is a $\delta$ Scuti pulsator and one of the most rapidly rotating stars in the solar neighborhood \citep[e.g.][]{vanBelle01,Buzasi05}. It rotates at $\sim$74\% of its angular (breakup) velocity with an equatorial velocity likely exceeding 300 km\,s$^{-1}$ \citep{Rieutord23}. This extreme spin rate flattens the star: interferometric observations directly resolved this oblateness, finding the equatorial radius $\sim$22\% larger than the polar radius, and produced the first resolved image of Altair, revealing strong gravity darkening \citep{Monnier07}. A 2D structural model using the ESTER code \citep{Bouchaud20} simultaneously fit these interferometric constraints alongside spectroscopic and asteroseismic data, yielding a mass of $1.86 \pm 0.03\ M_\odot$ and showed that the core rotates $\sim$50\% faster than the envelope. 

Most nearby, A-type stars (i.e., Vega and Fomalhaut) have been extensively studied for numerous imaging searches for planets from both ground and space based telescopes \citep{Heinze08,Janson15, Meshkat18, Ren23, Kalas08, Ygouf24, Beichman2025b}. Altair, however, has largely been overlooked from deep targeted analyses like Vega and Fomalhaut due to its assumed older age from 1D stellar evolution models (1.2--1.4 Gyr; \citealt{deSouza05}). A recent analysis of TESS data using 2D structural modeling \citep{Rieutord24} derived a younger age for Altair than previous estimates: $88\pm10$ Myr. Altair is now the closest A-star with an age under 100~Myr, making it an ideal and exciting new candidate for direct imaging searches for young giant exoplanets. JWST coronagraphy is particularly well suited for this search, as its deep infrared sensitivity enables access to sub-Jupiter mass companions around young stars \citep{Carter21,Carter23}, as recently demonstrated by the discovery of a sub-Jupiter mass planet TWA 7b \citep{Lagrange25,Crotts25}

Despite its youth, Altair shows no evidence for a debris disk. Photometric observations at 24~$\mu$m show no significant infrared excess \citep[1\%;][]{Rieke05}, and resolved imaging in the near-infrared and submillimeter also show no disk detection \citep{Thureau14, Spalding22}. The lack of a detected disk around Altair is surprising since its revised age ($88\pm10$ Myr) puts it comfortably in the range when debris disks around A-type stars are most commonly detected \citep[tens to hundreds of Myr;][]{Wyatt08}. Notably, near-infrared interferometry with CHARA/FLUOR does reveal a $3.07\pm0.24$\% $K$-band excess around Altair, attributed to an extended circumstellar source \citep{Absil13}, suggesting some circumstellar material may persist in the inner system.

\begin{deluxetable*}{llll}[thb!]
\tabletypesize{\scriptsize}
\tablewidth{0pt}
\tablecaption{Properties of the Altair \label{tab:altair_properties}}
\tablehead{
\colhead{Property} & \colhead{Value}& \colhead{Units}& \colhead{Comments} }
\startdata
Spectral Type & A7V & & \citet{Gray03}\\
Apparent $V_{mag}$ & $0.765\pm0.015$ & mag & \citet{Mermilliod97}\\
Apparent $M_{mag}$ & $0.245\pm0.025$ & mag & \citet{Bouchet91}\\
T$_{\rm pole}$ & 8450$\pm$140 & K & \citet{Monnier07}\\
T$_{\rm eq}$ & 6860$\pm$150 & K & \citet{Monnier07}\\
Mass & 1.86$^{+0.03}_{-0.03}$ & M$_\odot$& \citet{Bouchaud20}\\
log($L/L_{\odot}$) & 1.027$\pm$0.011 & dex & \citet{Peterson06}\\
Age$^a$ & 88$\pm10$ &Myr & \citet{Rieutord24}\\
$[$Fe/H$]$ & -0.2 & dex & \citet{Monnier07}\\
log($g$)& 4.266$\pm$0.012 &cgs & \citet{Peterson06}\\	 
inclination ($i$)& $50^{\circ}.65\pm1^{\circ}.23$ &&\citet{Bouchaud20}\\	
$\theta_{\rm pole}$ & 2.958$\pm$0.020 & mas & \citet{Monnier07}\\
$\theta_{\rm eq}$ & 3.670$\pm$0.014 & mas & \citet{Monnier07}\\
$\alpha_{ICRS}$ (epoch J2000) &19$^h$50$^m$47.00$^s$ & & \citet{vanLeeuwen07} \\
$\delta_{ICRS}$ (epoch J2000) &$+$08$^{\circ}$52$^\prime$05.96\arcsec\ & & \citet{vanLeeuwen07} \\
$\alpha_{ICRS}$ (epoch J2024.78) &19$^h$50$^m$47.88$^s$ & & This work \\
$\delta_{ICRS}$ (epoch J2024.78) &$+$8$^{\circ}$52$^\prime$15.09\arcsec\ &  &This work  \\
$\alpha_{ICRS}$ (epoch J2025.40) &19$^h$50$^m$47.92$^s$ &  & This work  \\
$\delta_{ICRS}$ (epoch J2025.40) &$+$8$^{\circ}$52$^\prime$16.18\arcsec\ &  & This work  \\
Parallax ($\varpi$) & $194.95\pm0.57$ & mas & \citet{vanLeeuwen07}\\
Distance ($d$) & 5.130$\pm$0.015 & pc & \citet{vanLeeuwen07}*\\
$\mu_{\alpha}$ & $536.024\pm0.0420$ & \masyr & \citet{Munn22}\\
$\mu_{\delta}$ & $385.173\pm0.0560$ & \masyr & \citet{Munn22}\\
$U$ & $-29.15\pm0.31$ & \kms & this work\\
$V$ & $-10.43\pm0.19$ & \kms & this work\\
$W$ & $-2.55\pm0.18$ & \kms & this work\\
\enddata
\tablecomments{$^*$As observed from vantage point of JWST at the epoch of observation.
* Distance $d$ calculated as 1/$\varpi$. Galactic Cartesian $U, V, W$ velocity vectors in barycentric frame, with $U$ towards Galactic center,
$V$ in direction of Galactic rotation, $W$ towards North Galactic Pole, calculated using {\it astropy} {\it SkyCoord}.}
\end{deluxetable*}

Altair's brightness, lack of a debris disk, and absence of known companions also make it a promising candidate as a reference star for future high-contrast coronagraphic imaging campaigns, both from the ground and from space potentially with Roman Coronagraph Instrument \citep{Bailey23,Hom26}, further motivating a thorough characterization of its circumstellar environment in this work with JWST/NIRCam.

\section{Observations} \label{Sec:obs}

We observed Altair with JWST/NIRCam on 2024 October 11 as part of the NIRCam GTO program (PI. Beichman, Program ID \#4534)\footnote{All the JWST data used in this paper can be found in MAST: \dataset[10.17909/61yw-dq54]{https://doi.org/10.17909/61yw-dq54}.}. Altair was observed simultaneously in the long (F444W, $4.44\;\mu$m) and short (F200W, $2.0\;\mu$m) wavelength channels using the M430R round Lyot coronagraph, which has an inner working angle of $0\farcs43$ in subarray mode ($\pm$10\arcsec\ field of view). This choice of mask minimizes the effects of saturation at the modest expense of inner angle ($<$1\arcsec). We observed Altair at two roll angles separated by $\sim$8 $^\circ$ to enable Angular Differential Imaging (ADI; \citet{Marois06, Liu04}). Each roll consisted of 650 integrations of 5 groups in RAPID readout mode, for a total integration time of 4182 s per roll and a total on sky integration time of 8364 s ($\sim$2.3 hr) after combining both rolls. We also observed Vega as a reference star for perform Reference Differential Imaging (RDI) using the same filter and mask configuration, with 5 dithered pointings of 140 integrations each (700 total integrations). Both targets were acquired using the F335M filter with the neutral density square. \autoref{tab:obs} summarizes the observing parameters. 

By observing Altair in both F444W and F200W, we were able to reject out any detected sources as background stars, since stars are significantly brighter in 2 $\mu$m than in 4.44 $\mu$m. However, since the M430R mask only covers one half of the sky in F200W, we only can rule out detected companions to the south of F444W (see \autoref{fig:altair_images}). 

We obtained second epoch observations of Altair with NIRCam on 2025 May 26 (PI Meshkat, Program ID \#8507). We used the same filters, roll angles, reference star, masks and integration times, in order to have as comparable observations as possible, with the primary difference that the observation was scheduled at a time where the M430R mask covered the sky to the North of Altair for the F200W images. 

\begin{deluxetable*}{lllllll}
\tabletypesize{\scriptsize}
\tablewidth{0pt}
\tablecaption{NIRCam Coronagraphic Imaging Observing Parameters\label{tab:obs}}
\tablehead{
\colhead{Target} & \colhead{Filter} & \colhead{Readout} & \colhead{Groups/Int} & \colhead{Ints/Exp} & \colhead{Dithers} & \colhead{Total Time (s)}}
\startdata
\multicolumn{7}{c}{{Epoch 1 - 2024 Oct 11}} \\ \hline
Altair (Roll 1) & F200W/MASK430R & SUB320/RAPID & 5 & 650 & 1 & 4182 \\
Altair (Roll 1) & F444W/MASK430R & SUB320/RAPID & 5 & 650 & 1 & 4182 \\ \hline
Altair (Roll 2) & F200W/MASK430R & SUB320/RAPID & 5 & 650 & 1 & 4182 \\
Altair (Roll 2) & F444W/MASK430R & SUB320/RAPID & 5 & 650 & 1 & 4182 \\ \hline
Vega           & F200W/MASK430R & SUB320/RAPID & 5 & 140 & 5 & 901 \\
Vega           & F444W/MASK430R & SUB320/RAPID & 5 & 140 & 5 & 901 \\ \hline\hline
\multicolumn{7}{c}{{Epoch 2 - 2025 May 26}} \\ \hline
Altair (Roll 1) & F200W/MASK430R & SUB320/RAPID & 5 & 650 & 1 & 4182 \\
Altair (Roll 1) & F444W/MASK430R & SUB320/RAPID & 5 & 650 & 1 & 4182 \\ \hline
Altair (Roll 2) & F200W/MASK430R & SUB320/RAPID & 5 & 650 & 1 & 4182 \\
Altair (Roll 2) & F444W/MASK430R & SUB320/RAPID & 5 & 650 & 1 & 4182 \\ \hline
Vega          & F200W/MASK430R & SUB320/RAPID & 5 & 140 & 5 & 901 \\
Vega          & F444W/MASK430R & SUB320/RAPID & 5 & 140 & 5 & 901 \\
\enddata
\tablecomments{Roll angle separation between Roll 1 and Roll 2 was $8.61\degr$ (Epoch 1) and $8.00\degr$ (Epoch 2). Both targets were acquired using F335M with the neutral density square. Vega served as the PSF reference star for RDI in both epochs, with 5 dithers for PSF diversity.}
\end{deluxetable*}

\section{Data Reduction and Post-Processing} \label{Sec:data_reduction}
\subsection{Data Reduction}

Basic data reduction is done following \citet{Ygouf24}.
For each telescope pointing, the images are flat-fielded, background subtracted, flagged for cosmic rays, co-added over multiple exposures, flux calibrated, and distortion corrected. At this level-2 stage of processing, images at the two telescope roll angles have not yet been combined, nor has the reference star been subtracted.

The data was processed with pipeline version \rm{2025\_2},
calibration version \rm{1.9.6},
CRDS context for reference files \rm{jwst\_1413.pmap}, and
distortion reference file \rm{jwst\_nircam\_distortion\_0173.asdf}. 
The dataset can be obtained at: \dataset[https://doi.org/10.17909/61yw-dq54]{https://doi.org/10.17909/61yw-dq54}.

We have made some adjustments to the standard pipeline, to remove instrumental artifacts such as bad pixels and striping resulting from 1/f noise. While the standard pipeline flags pixels neighboring saturation-flagged pixels, we do not see any evidence of charge spillage and have removed those flags. {\bf We use the overarching ``DO\_NOT\_USE" pipeline flag to identify and NaN out other bad pixels (e.g.\ dead pixels, hot pixels, those without a linearity correction, etc.).}

As the final pre-processing step, each data cube of 650 individual frames was collapsed to a single median frame prior to PSF subtraction.
Before this data stacking we remove pixels that are 5-$\sigma$ outliers temporally (an outlier within one of the 650 sub-exposures) and then we remove outliers spatially (an outlier within a 5x5 box of surrounding pixels, after co-adding the sub-exposures).

\subsection{Post-Processing}

\begin{figure*}[htbp]
\centering
\includegraphics[width=0.48\textwidth]{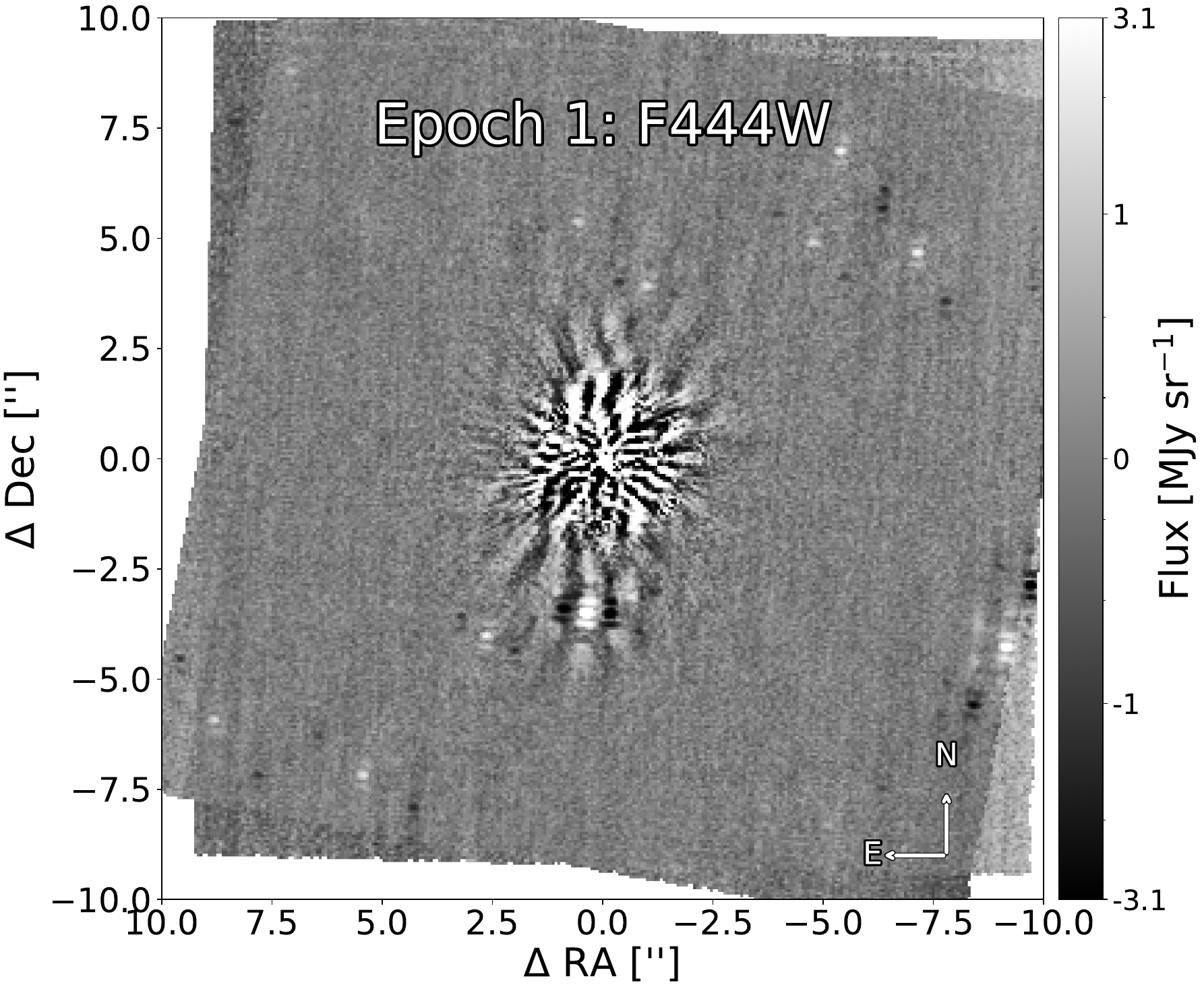}
\includegraphics[width=0.48\textwidth]{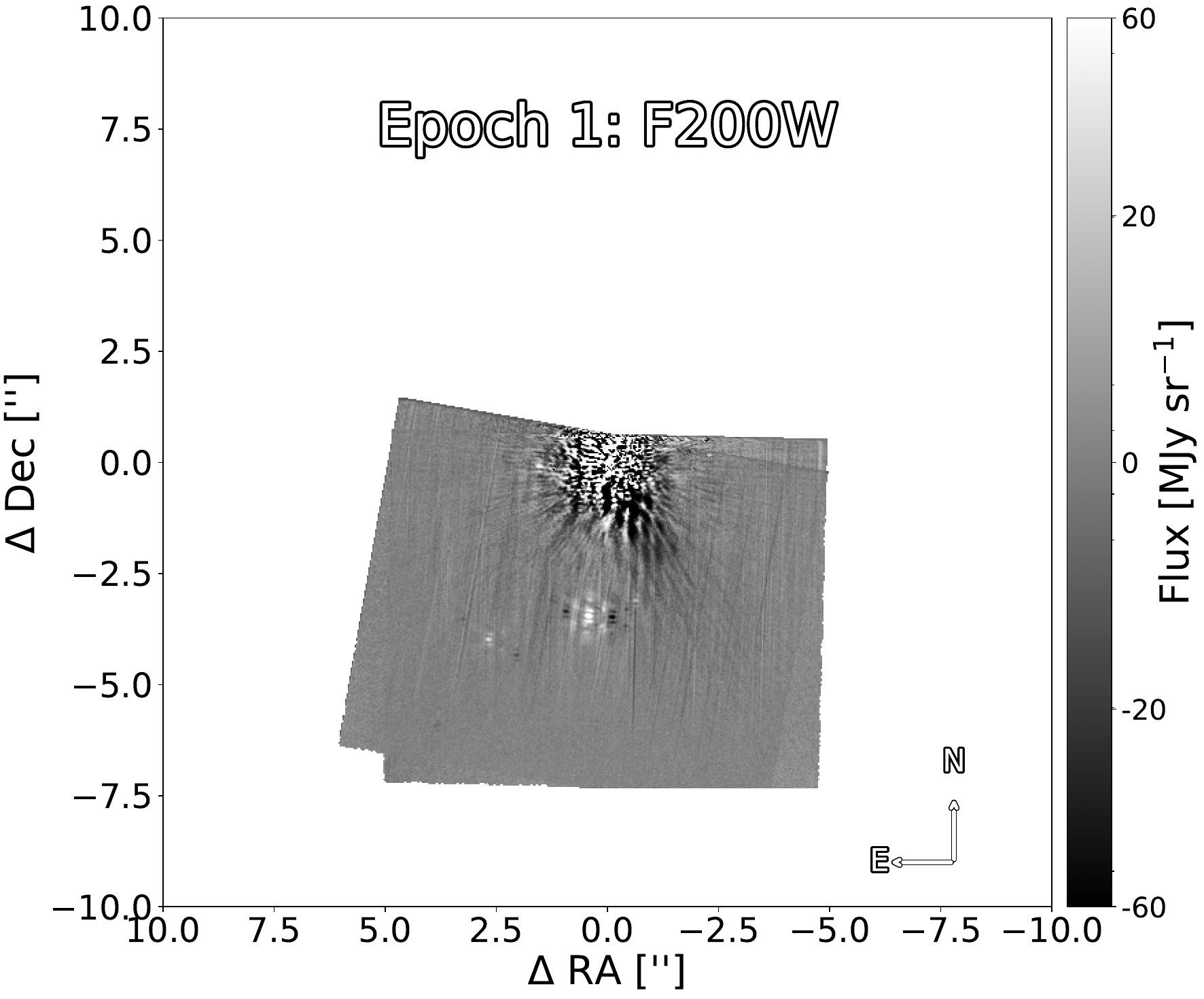}\\
\includegraphics[width=0.48\textwidth]{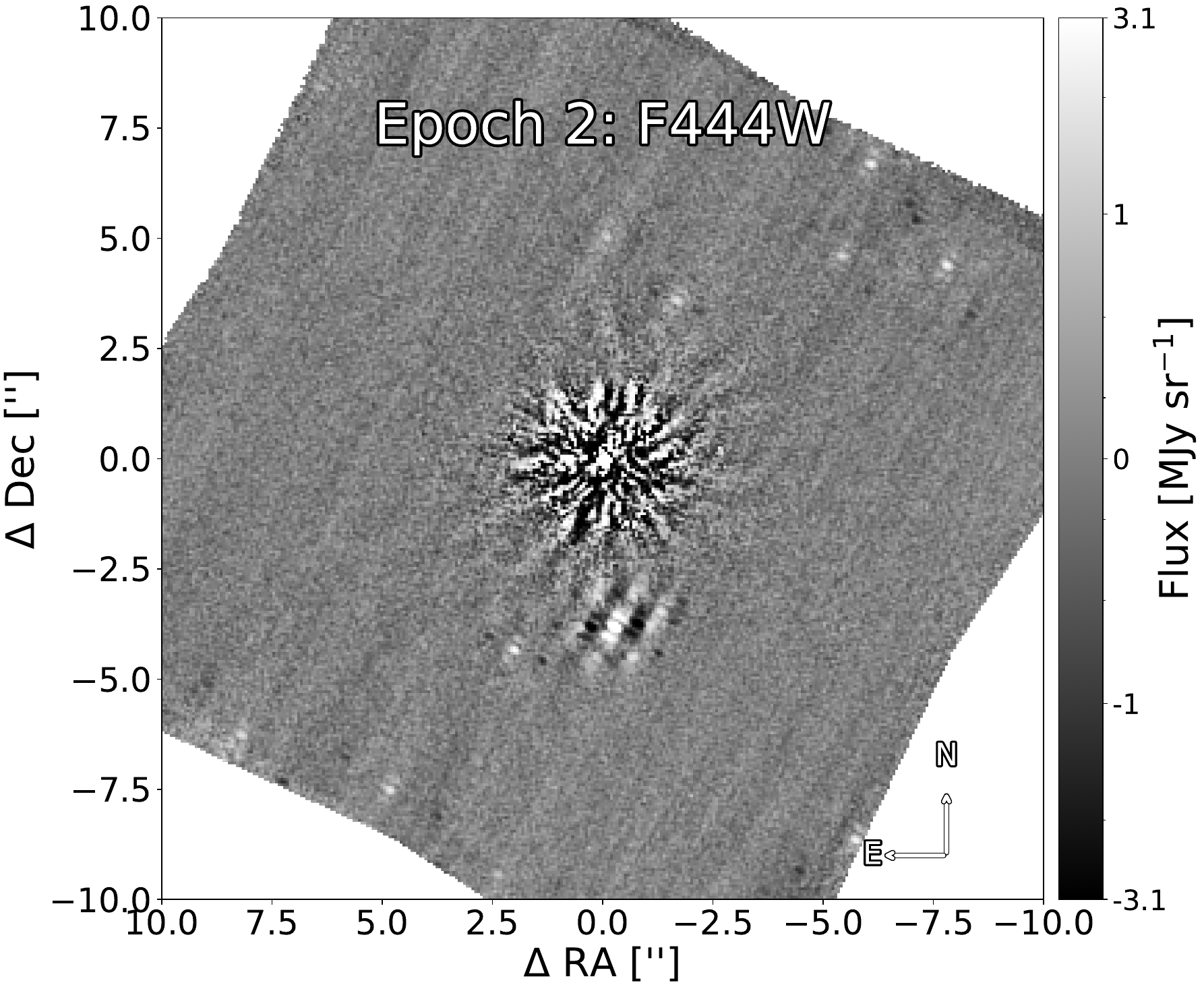}
\includegraphics[width=0.48\textwidth]{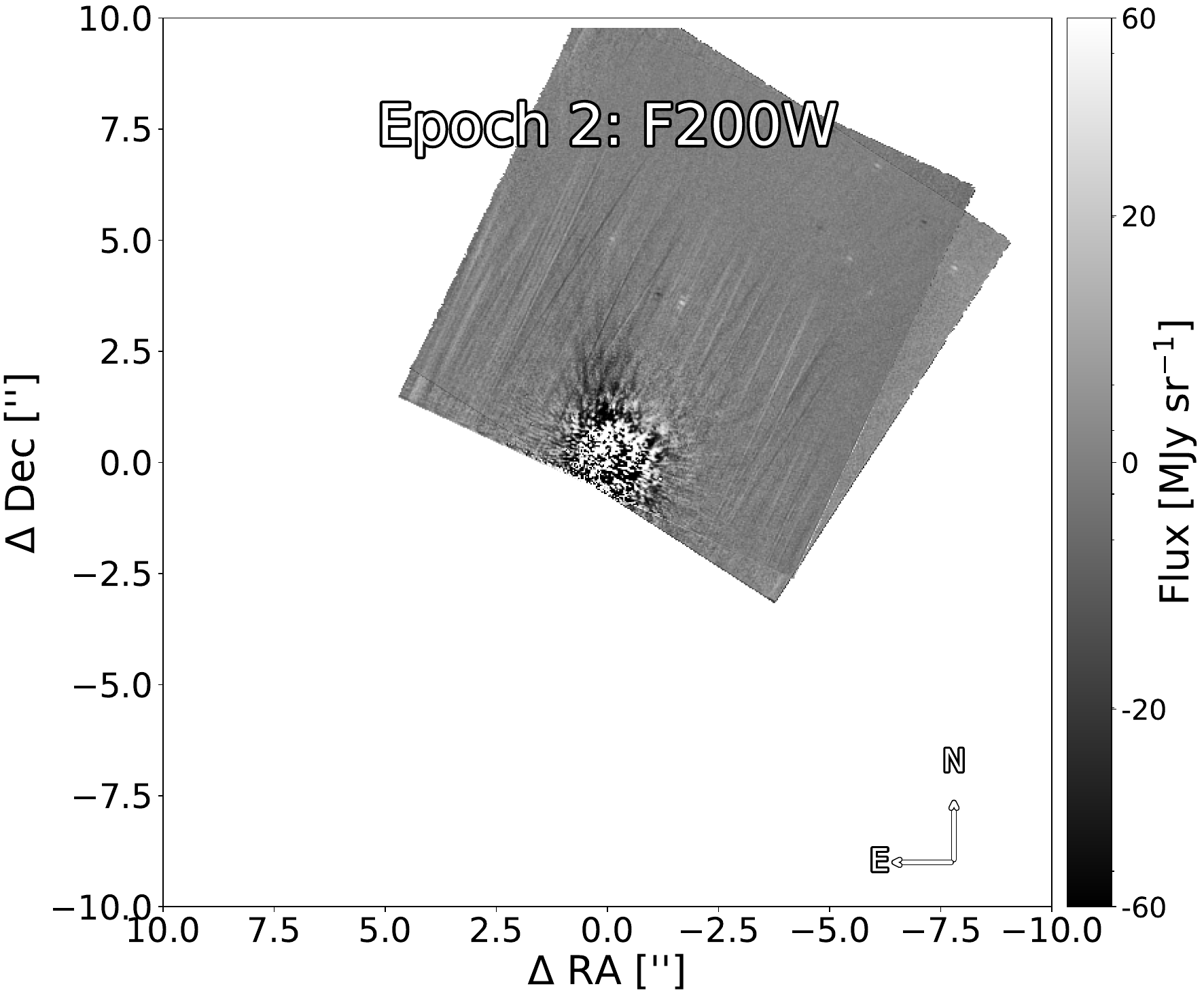}
\caption{Post-processed images of Altair (ADI+RDI, 5 KL modes) with NIRCam. The top row shows Epoch 1 in F444W (left) and F200W (right); the bottom row shows Epoch 2 in the same filters. North is up, east is to the left. Point sources around Altair were shown to be background stars using a combination of F200W-F444W colors and lack of common proper motion with Altair in the second Epoch data. The negative sources on the images are due to the ADI subtraction.}
\label{fig:altair_images}
\end{figure*}

Point-spread function (PSF) subtraction was performed using the KLIP algorithm \citep{Soummer12} using the spaceKLIP (v2.2.1; \citealt{Kammerer22, Carter23, Girard22}), which is a community supported python package designed for data reduction and analysis of JWST high-contrast imaging data. For the PSF subtraction algorithm, we used the pyKLIP \citep{Wang15} implementation of KLIP. Both epochs of our NIRCam data (PID 4534 and 8507) were reduced identically in our pre and post-processing pipelines. We performed the PSF subtraction using ADI \citep{Marois06, Liu04} with the median coadded frames at two rolls, RDI \citep{Lafreniere09} using Vega as the reference star, and ADI+RDI which uses both. We chose to do the PSF subtraction with a single annulus and subsection (meaning full-frame subtraction). We performed the PSF subtraction for several KL modes as is standard in spaceKLIP including 1, 2, 5, and 10 modes. \autoref{fig:altair_images} shows the final post-processed images of Altair with 5 KL modes at both F444W and F200W, for the two epochs.

\section{Results} \label{Sec:analysis}
\subsection{Sensitivity limits}

\begin{figure*}[htbp]
\centering
\includegraphics[width=0.48\textwidth, trim=0.7cm 0cm 0.0cm 0cm, clip]{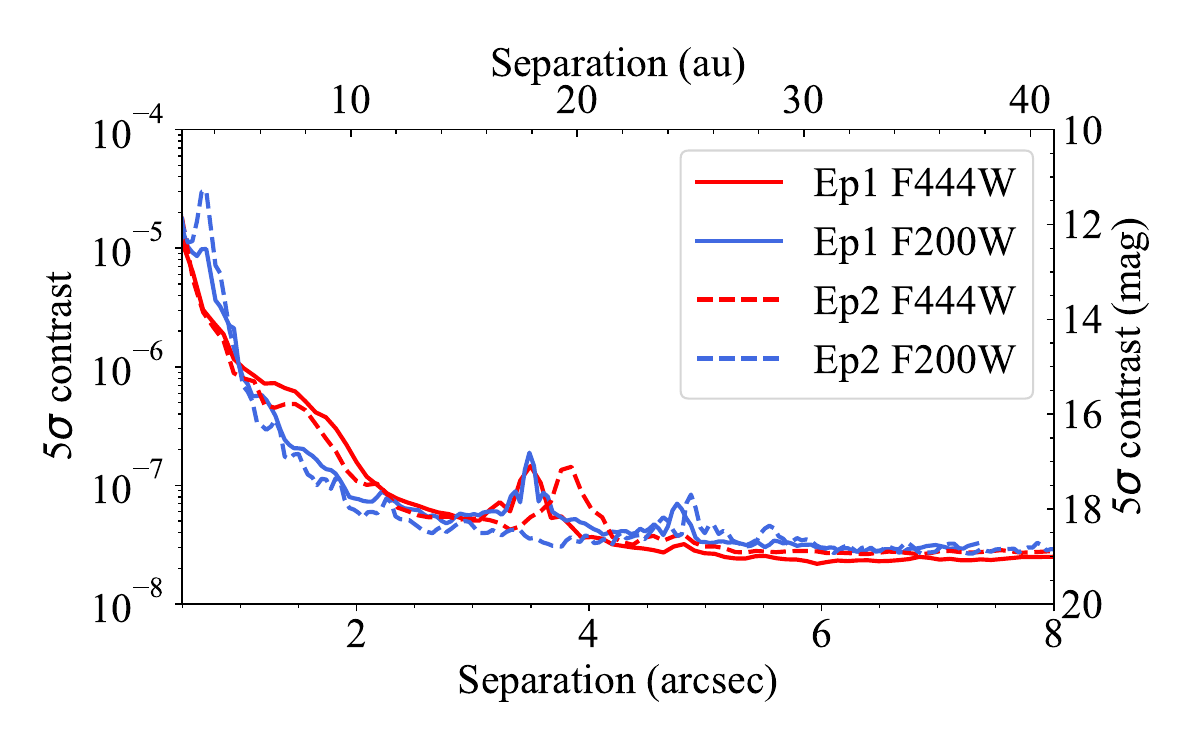}
\includegraphics[width=0.48\textwidth, trim=0.7cm 0cm 0.9cm 0cm, clip]{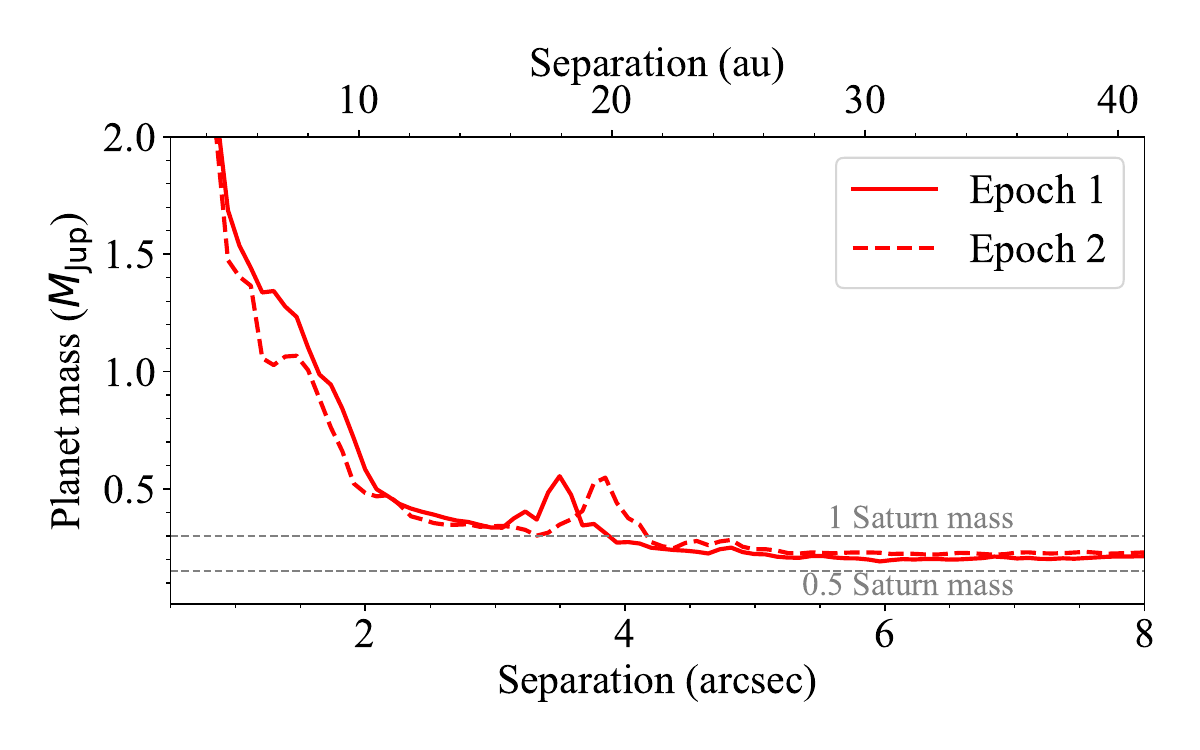}
\caption{Sensitivity achieved in the JWST/NIRCam Altair data, showing deep limits on solar system orbital scales. \textit{Left:} Contrast limits for the first epoch F444W JWST/NIRCam Altair data. The bump in contrast around 3.5” corresponds to the bright star south of Altair, visible in \autoref{fig:altair_images}. \textit{Right:} Mass limits for the F444W NIRCam data computed using the [M/H] = +0.5~dex metallicity, solar C/O ratio, $\log K_{\rm zz} = 7$~dex Sonora Flame Skimmer model \citep{Mang2026}, showing sensitivities down to 0.7 Saturn mass.}
\label{fig:contrast}
\end{figure*}

Contrast curves were computed using the spaceKLIP \texttt{AnalysisTools} module from the ADI+RDI PSF-subtracted images for 5 KL modes. Raw 5$\sigma$ contrast curves were first calculated by computing the noise in azimuthal annuli at each separation and normalizing to the synthetic peak flux of the stellar PSF, estimated from a Vizier VOTable of Altair photometry\footnote{Altair data were retrieved from Vizier on 2024-10-12, with a search radius of 1''. One photometric point was excluded (Johnson:K, \citealt{Morel78}) due to anomalously large error bars.} fit to a spectral type of A7V using \texttt{STPSF}. The raw contrast was then calibrated via fake companion injection-recovery tests to account for KLIP algorithmic throughput losses, with companions injected at separations of $0\farcs5$, 1$''$, 2$''$, 3$''$, and 4$''$ and position angles of 45$\degr$ and 90$\degr$ at a flux level of 20$\sigma$. We additionally corrected for small sample statistics at the closest separations \citep{Mawet14} and accounted for coronagraphic mask throughput losses. Calibrated contrast curves for both epochs in F444W and F200W are presented in \autoref{fig:contrast} (left).

\autoref{fig:contrast} (right) converts the F444W curves to planet mass limits using the Sonora Flame Skimmer (cloud-free) atmospheric and evolutionary models \citep{Mang2026}, which extend to sub-Saturn masses whereas many first-generation models (e.g., COND; \citealt{Baraffe03}) reach their lower boundary at $\sim$0.5~$M_\mathrm{Jup}$. Specifically, we use the [M/H] = +0.5~dex metallicity, solar C/O ratio, $\log K_{\rm zz} = 7$~dex models as a representative planet scenario. Note that mass limits are model-dependent and alternate atmospheric scenarios are discussed in \autoref{sec:appendix_mass_limits}. We adopt Altair's age, distance, and F444W magnitude from \autoref{tab:altair_properties} \citep{Rieutord24, vanLeeuwen07}. We do not convert F200W contrast to planet mass as it is primarily used for background star discrimination, as young planets emit negligibly at 2 $\mu$m relative to an A-type host. Interior to $\sim$2$''$ ($\sim$10~au), sensitivity is limited by coronagraphic mask throughput and residual PSF subtraction artifacts. In the background-limited regime beyond this radius, we achieve a maximum contrast of $\sim$19~mag, corresponding to $\sim$0.7~$M_{\rm Sat}$. We find no companion candidates with common proper motion with Altair, placing meaningful constraints on the giant planet population around Altair and making this one of the most sensitive direct imaging searches for low-mass companions around a nearby A-type star\footnote{\citet{Beichman2025b} achieved similar contrast around Vega with JWST/NIRCam data, however the mass limits presented in this work are deeper due to the younger age of Altair.}.

\subsection{Point Sources Detected near Altair}

In the first-epoch F444W images (\autoref{fig:altair_images}, top left), we identified many point sources within $10''$ of Altair by visual inspection. Sources to the South were quickly confirmed to be background stars based on their brightness in F200W (where planets are expected to be fainter) and F200W-F444W color. The selected mask (M430R) limits the field of view in the F200W image to the southern half of Altair, leaving any Northern point sources as companion candidates. These northern candidates lie at separations of $\sim4''$ and have contrasts of $\sim$17.4 mag relative to Altair. If gravitationally bound, they would correspond to Saturn-mass companions or less at projected separations of 20-30 au. In our second-epoch observations, taken $\sim$7 months later, we were able to show that the candidate companions to the North are background stars, due to their lack of common proper motion with Altair, and their F200W-F444W colors (\autoref{fig:altair_images}, bottom).

\subsection{Bright Background Stars}
\label{subsec:bright_background_stars}

\begin{table*}[htbp]
\centering
\caption{Astrometry and photometry of the brightest background star (ID 1) south of Altair \label{tab:bgstar}}
\begin{tabular}{clcccccl}
\hline\hline
ID & Date & Filter & $\Delta$RA ($''$) & $\Delta$Dec ($''$) & PA ($\degr$) & Sep. ($''$) & Contrast \\
\hline
\multirow{3}{*}{1}
    & 2024 Oct 11 & F200W & $+0.431\pm0.001$ & $-3.468\pm0.001$ & $172.9\pm0.01$ & $3.495\pm0.001$ & $3.92\pm0.03 \times 10^{-7}$\\
    & 2024 Oct 11 & F444W & $+0.398\pm0.001$ & $-3.485\pm0.001$ & $173.5\pm0.02$ & $3.507\pm0.001$ & $4.96\pm0.04 \times 10^{-7}$  \\
    & 2025 May 26 & F444W & $-0.226\pm0.001$ & $-3.807\pm0.001$ & $183.4\pm0.02$ & $3.814\pm0.001$ & $4.89\pm0.04 \times 10^{-7}$\\
\hline
\end{tabular}
\begin{minipage}{\textwidth}
\smallskip
\textit{Note.} Astrometry and contrast measured from JWST/NIRCam coronagraphic imaging using \texttt{spaceKLIP}. Positive $\Delta$RA is East, positive $\Delta$Dec is North. The source is confirmed as a background star based on its motion relative to Altair between epochs and its F200W-F444W color (\autoref{tab:companion_colors}). For the 10 less bright candidate sources identified in the field, see \autoref{fig:companion_finder} and Table~\ref{tab:companions} in the Appendix.
\label{tab:bgstar}
\end{minipage}
\end{table*}

As Altair has no detected companions and no cold debris disk \citep{Thureau14, Spalding22}, it is a promising candidate reference star for future high-contrast coronagraphic imaging campaigns from the ground or space. We therefore present the position and brightness of 11 background stars in our image to support future observation planning. The source properties measured of the brighest background star due south of Altair data are summarized in \autoref{tab:bgstar}, the other ten fainter objects identified in the field are described in \autoref{fig:companion_finder} and Table~\ref{tab:companions} in the Appendix. The astrometry and photometry for all point sources were measured using spaceKLIP's \textit{extract\_companions} function, which forward-models the JWST PSF and jointly fits position and flux via MCMC (Table~\ref{tab:bgstar} and \ref{tab:companions}).

We note that Altair is so nearby that its large angular diameter ($\sim3$--$3.7$ mas; \citealt{Monnier07,Bourges17,vanBelle01}) may result in it being partially resolved by some telescopes. This angular diameter exceeds the $<2$ mas ideal Roman Coronagraph Instrument reference star diameter \citep{Hom26}. However, as the best reference stars for Roman are still being determined, we present this information for the Roman Coronagraph community and any other future high-contrast imaging campaigns where partial resolution of the stellar photosphere is not a limiting factor.

In the first epoch data, we identified a bright point source located almost due south of Altair at a separation of $\sim$3.5$''$. Due to the high proper motion of Altair (\autoref{tab:altair_properties}), our second epoch observations just $\sim$7 months later show the bright source moves relative to Altair by $\sim$700 mas, confirming it is a stationary background star via proper motion. This is further supported by its F200W-F444W color ($+0.27\pm0.07$; Table~\ref{tab:companion_colors}), which is consistent with a background star rather than the much redder colors expected of a young, self-luminous planet. Additionally, the point sources we identified in the field of view (Table~\ref{tab:companions}) and show that they all have consistent apparent motion relative to Altair (Table~\ref{tab:companion_colors}), both in on sky motion ($\sim$0.64--0.76$''$) and direction, further confirming their status as background stars rather than physically associated companions.

We additionally note that, looking ahead to future coronagraphic imaging campaigns which may use Altair as a reference star, the sky background in the vicinity of Altair will be relatively uncontaminated by background stars following the passage of this bright source (see \autoref{fig:altair_images}). Of the point sources identified in this work (Table~\ref{tab:companions}), only objects 7 and 11 lie in the direction of Altair's proper motion. The other sources will continue to become more distant in projected separation. Object 7 will make its closest approach at $\sim$4.1$''$ separation around late 2029 (contrast $\sim$7.4$\times10^{-8}$ in F444W), and object 11 at $\sim$3.1$''$ around late 2040 (contrast $\sim$5.8$\times10^{-8}$ in F444W). Both are about an order of magnitude fainter than the bright background star characterized in Table~\ref{tab:bgstar}, and neither will pass closer than $\sim$3.1$''$, suggesting a particularly favorable window for Altair as a reference star for high contrast imaging in the coming years.

\section{Discussion} \label{sec:Discussion}

Our deep NIRCam observations, sensitive to sub-Saturn-mass planets beyond $2''$, reveal no confirmed companions and earlier observations revealed no debris disks emission. This contrasts with other A stars like Vega and $\beta$ Pic which host either a disk \citep{Smith84,Aumann84}, planets \citep{Lagrange10} or both. Based on planet occurrence rates and mass functions presented in \citet{Meyer25}, even with a null detection in our deep dataset, $< 1-2 M_{\rm Jup}$ planets may still reside within 10 au where occurrence rates are expected to be higher.

Altair is somewhat unusual but not  exceptional in its lack of a disk and its high rotational velocity (v sin i =221 km s$^{-1}$ \citep{Erspamer2003}. The simplest explanation for  these properties is that Altair is one of those A stars born with a high rotational velocity and  which has not been appreciably slowed given its relatively youth and the inefficiency of magnetic braking in hot stars with thin or non-existent convective layers \citep{Zorec2012}. 

A more exotic explanation of the high rotational velocity is that Altair ingested a massive planet or low mass star with a consequent increase in its rotational angular momentum. The presence of such a companion on an eccentric orbit and eventually fatal trajectory could also clean out a residual debris disk and account for Altair's lack of a significant level of IR emission \citep{Beichman2025}. Ingestion of a companion, particularly a planetary mass object, would be expected to change the star's elemental abundances \citep{Behmard2022}. Measuring elemental abundances for Altair is challenging given its rapid rotation ($v\sin i \sim 240$ km s$^{-1}$). The recent spectroscopic analysis by \citet{Takeda2026} measuring the abundances of 17 elements finds that Altair is slightly subsolar ([Fe/H] $\sim -0.2$), differential abundances with respect to the Sun are consistent with $-0.5 < \mathrm{[X/Fe]} < +0.3$, lacking any signs of a dependence on atomic number, and no similarity to other peculiar classes (e.g. Am, lamba Boo stars). Hence, Altair appears to be a ``normal” A-type star, not showing any abundance effects that might indicate evidence of past planetary engulfment. 

Finally, while debris disks among young A stars are common, there are numerous examples of young A stars without detectable disks \citep{su2006}. The lack of any planet could be attributed to rapid dissipation of the protostellar disk \citep{Ribas2015}.

\subsection{A Search for Altair Siblings}\label{sec:siblings}

As there is some disagreement in the literature on the age of Altair, and given the surprise of the extremely young age of $88\pm10$ Myr recently derived by \citet{Rieutord24}, we searched for any evidence of low-mass stellar companions or a moving group to which Altair might belong. 

\subsubsection{Search for a Kinematic Group}

We calculate an updated space velocity for Altair based on combining the available astrometry and radial velocity estimates with the smallest uncertainties, including parallax from \citet{vanLeeuwen07}, proper motion and ICRS position from \citet{Munn22}, and radial velocity from \citet{Gontcharov06}. Based on these values, we estimate the Galactic Cartesian velocity of Altair (in the solar system barycentric frame) using {\it Astropy} {\it SkyCoord} to be ($U, V, W$) = ($-29.15, -10.43, -2.54$) $\pm$ ($0.27, 0.29, 0.06$) \kms.

The BANYAN $\Sigma$ (Bayesian Analysis for Nearby Young AssociatioNs $\Sigma$) tool from \citet{Gagne18} was used to search for potential host stellar association for Altair. The SIMBAD default values for astrometry from \citet{vanLeeuwen07} and the literature mean radial velocity from \citet{Gontcharov06} were taken as inputs. BANYAN $\Sigma$ estimated the probability of the star being ``field'' was 99.7\%\, and attributed a 0.3\%\, probability of membership to the 40 Myr-old Argus group, and a $<$0.1\% membership probably to the other $d$ $<$ 150\,pc stellar groups surveyed by \citet{Gagne18}.

\citet{Gagne26} updated the BANYAN $\Sigma$ model to included recent Gaia results related to the discovery of new kinematic groups and membership analyses, creating the Montreal Open Clusters and Associations database (MOCAdb). A query of MOCAdb using the astrometry from \citet{vanLeeuwen07} and radial velocity from \citet{Kharchenko07}, yielded a 98.9\%\ probability of Altair being a field star, with negligible probabilities that it could belong to Carina-Near ("CARN"; 0.75\%) and [MGC2022] Crius 209 (0.33\%). We discuss these two groups further to see whether there is any plausibility of an genetic tie with Altair.

The Carina-Near group is a $\sim200$ Myr-old small group discovered by \citet{Zuckerman06} situated $\sim$30\,pc from the Sun. \citet{Mamajek16} notes that group is very small, with no known members hotter than F1, an estimated total mass of only $\sim8$\,\Msun, and an intrinsic 1D velocity dispersion of $\sim$1.3 \kms\, \citep{Mamajek16}. The velocity of Car-Near ($U,V,W$ = $-25.3, -18.1, -2.3$ \kms) differs from Altair by $\Delta S$ = 8.6\,\kms, at roughly $\sim$5.8$\sigma$ significance if one considers the 1D velocity dispersion of Car-Near, uncertainties in mean $UVW$ vector for Car-Near, and uncertainties in Altair's $UVW$. Altair is 32\,pc from the Car-Near centroid \citep{Gagne18}, and running the orbits of Altair and Carina-Near back in time using epicyclic approximation following \citet{Faramaz21}, we find that Altair was no closer than $\sim$15 pc from the Carina-Near centroid (3.4 Myr ago), and passed with relative velocity $\delta S$ $\simeq$ 8.3 \kms. Given the lack of more massive systems in the Carina-Near nucleus, an ejection would have required a more massive star ($\sim 4.2$ \Msun, if it ended its life 3 Myr ago within a $\sim$200 Myr-old cluster). However Altair shows no signs of anomalous elemental abundances that might suggest ejection from a close-in binary with a more massive partner.

\citet{Moranta22} first identified the candidate group [MGC2022] Crius 209 in their Gaia EDR3 6D clustering analysis using HDBSCAN, and presented a membership list. However the group failed two of their four criteria for declaring a new association for which they would quote an age, instead Crius 209 had notably large scatter in $Z$ and $W$. \citet{Moranta22} notes "{\it that these rejected Crius groups may contain additional coeval structures, but they will likely also include a larger number of contaminated or unphysical clusters}." The Crius 209 sample from \citet{Moranta22} does not include any stars brighter than $V$ or $G$ of 7th magnitude, and their distances are $d$ $\simeq$ $66\pm21$ pc (rms). 
Taking the Crius 209 centroid position and velocity from \citet{Moranta22} at face value, the velocity of Crius 209 differs from that of Altair by $\Delta S = 7.1$ \kms\ ($\chi^2/\nu$ = 8.6/3, $P$=0.036). 
Running the orbits of Altair and Crius 209 back in time using epicyclic approximation following \citet{Faramaz21}, we find that Altair came no closer than about 32\,pc from the Crius 209 centroid, and had relative velocity $\Delta S = 6.8$ \kms.
Again, there seems to be no smoking gun explanation for why
a late A star like Altair would be ejected from such a small group, itself containing only one other A-type star, and the physicality of the group itself is in question.

Hence, we conclude that it seems exceedingly unlikely that Altair could have any genetic connection to either Carina-Near or the putative Crius 209 group.

\subsubsection{Washington Double Star Catalog}

Stellar companions have proved valuable for constraining the ages of some of the nearest A-type stars \citep[e.g., Sirius, Fomalhaut;][]{Bond2017,Mamajek2012}. Given that Altair lies at only 5\,pc, it would have been remarkable indeed if a resolved stellar companion had been missed from recent previous surveys of nearby stars and wide companions \citep[e.g.][]{Kirkpatrick24,GonzalezPayo25}.

The Washington Double Star Catalog \citep{Mason01}\footnote{\url{https://astro.gsu.edu/wds/}} has six double star entries for Altair (WDS 19508+0852). In \autoref{tab:altair_WDS} we summarize the WDS entries for Altair (component A). Components B and C were already cross-IDed in SIMBAD. Using the WDS position angle and separations, we cross-IDed components D through G which were missing in SIMBAD. In \autoref{tab:altair_WDS2} we list the Gaia IDs, parallaxes, and proper motions for the WDS companions, and compare their values to those of Altair. The companions WDS 19508+0852B through G all have parallaxes and proper motions consistent with being distant ($>0.4$ kpc) background stars, and we conclude that none of the WDS entries associated with Altair are physical companions.

\begin{deluxetable}{lcrrr}[thb!]
\tabletypesize{\scriptsize}
\tablewidth{0pt}
\tablecaption{WDS Double Stars Associated with Altair (WDS J19508+0852)\label{tab:altair_WDS}}
\tablehead{
\colhead{Disc.} & \colhead{Comps.} & \colhead{$\theta$}& \colhead{$\rho$} & \colhead{mag}\\
\colhead{ID} & \colhead{...} & \colhead{$\circ$} & \colhead{\arcsec} & \colhead{...}}
\startdata
STFB 10	& AB & 286 & 195.80 &  9.82 \\
STFB 10	& AC & 110 & 186.40	& 10.30 \\
DAL 27	& AD & 105 &  26.80	& 11.90 \\
SMR 5	& AE & 351 & 151.70	& 11.00 \\
SMR 5	& AF & 48  & 292.40	& 10.30 \\
GMC 5	& AG & 121 & 185.10	& 13.00 
\enddata
\tablecomments{Columns:
(1) WDS discoverer ID, 
(2) component letters for WDS 19508+0852 (A = Altair),
(3) position angle $\theta$ ($\circ$),
(4) separation $\rho$ (\arcsec),
(5) apparent magnitude in WDS (system not specified).}
\end{deluxetable}

\begin{deluxetable*}{lccrrl}[thb!]
\tabletypesize{\scriptsize}
\tablewidth{0pt}
\tablecaption{Cross-IDs for Altair's WDS entries\label{tab:altair_WDS2}}
\tablehead{
\colhead{WDS ID} & \colhead{Alias} & \colhead{$\varpi$} & \colhead{\pmra} & \colhead{\pmdec} & \colhead{Ref.}\\
\colhead{...} & \colhead{...} & \colhead{(mas)} & \colhead{\masyr} & \colhead{\masyr} & \colhead{...}}
\startdata
WDS J19508+0852A & Altair                       & $194.95\pm0.57$   & $536.024\pm0.042$ & $385.173\pm0.056$ & 1,2\\
WDS J19508+0852B & Gaia DR3 4298633621839700736 & $2.0429\pm0.0132$ & $7.230\pm0.012$ 	 & $-1.974\pm0.010$ & 3\\
WDS J19508+0852C & Gaia DR3 4298620668218220160 & $1.0635\pm0.0175$ & $-12.371\pm0.014$  & $-4.639\pm0.013$ & 3\\
WDS J19508+0852D & Gaia DR3 4298621664638734592 & $0.4832\pm0.0166$ & $-0.334\pm0.013$ 	 & $-1.201\pm0.012$ & 3\\
WDS J19508+0852E & Gaia DR3 4298623520076617728 & $1.3878\pm0.0153$ & $-6.628\pm0.014$ 	 & $-22.367\pm0.012$ & 3\\
WDS J19508+0852F & Gaia DR3 4298623416997887232 & $1.5098\pm0.0225$ & $-1.353\pm0.018$ 	 & $-6.228\pm0.017$ & 3\\
WDS J19508+0852G & Gaia DR3 4298620633858473344 & $0.6359\pm0.0209$ & $1.891\pm0.017$ 	 & $1.631\pm0.015$ & 3\\
\enddata
\tablecomments{References: (1) \citet{vanLeeuwen07}, (2) \citet{Munn2022}, (3) \citet{GaiaDR3}.}
\end{deluxetable*}

Thus far, our analysis is consistent with the notion that Altair is a single star that does not belong to any known stellar association, a perhaps surprising result given its young asteroseismic age.

\section{Conclusions} \label{Sec:Conclusions}

We observed Altair with the JWST/NIRCam coronagraph in F444W and F200W, achieving contrast levels of $1.5 \times 10^{-7}$ at 2'' (10 au) and a few $\times 10^{-8}$ beyond 4'' (20 au). Using its recently revised age of 88$\pm$10 Myr \citep{Rieutord24} and the Sonora Flame Skimmer model \citep{Mang2026}, this corresponds to sub-Saturn mass limits beyond 10 au. We identified several companion candidates in the first epoch data. Second epoch observations showed these candidates were not co-moving with Altair, and their colors consistent with background stars. Altair remains elusive, with no detected debris disk, companions, and it is not known to be part of any stellar association. Its revised young age and proximity makes it a compelling target for future high-contrast imaging observations. We also characterize the background star field around Altair and find that it will remain a few arcsec from any background sources over the coming decade, making it a reliable reference star for future coronagraphic imaging campaigns, particularly for the Roman Coronagraph Instrument.

\begin{acknowledgments}
We thank the anonymous referee for helpful comments and
suggestions that improved this paper. This work was authored by employees of Caltech/IPAC under Contract No. 80GSFC21R0032 with the National Aeronautics and Space Administration.
Part of this research was carried out at the Jet Propulsion Laboratory, California Institute of Technology, under a contract with the National Aeronautics and Space Administration (80NM0018D0004). A.S acknowledges support from the National Science Foundation Graduate Research Fellowship under Grant No.~2139433.

This work is based on observations made with the NASA/ESA/CSA James Webb Space Telescope. The data were obtained from the Mikulski Archive for Space Telescopes at the Space Telescope Science Institute, which is operated by the Association of Universities for Research in Astronomy, Inc., under NASA contract NAS 5-03127 for JWST. These observations are associated with programs \#4534 and \#8507 and can be accessed via DOI:\dataset[https://doi.org/10.17909/61yw-dq54]{https://doi.org/10.17909/61yw-dq54}. The authors acknowledge the (NIRCam GTO team led by PI Beichman) for developing their observing program with a zero-exclusive-access period.

\end{acknowledgments}

\begin{contribution}

TM led the project, obtained the second epoch observations, performed the data analysis, and wrote and submitted the manuscript.
CB came up with the initial research concept, led the NIRCam GTO team, wrote the discussion of Altair's lack of a detected disk, and edited the manuscript.
JLS contributed to the post-processing analysis and edited the manuscript.
GB performed the data pre-processing and edited the manuscript.
AZG supported the data processing, provided expertise on NIRCam data analysis, and edited the manuscript.
EM led the search for Altair co-moving siblings and the stellar age analysis, and edited the manuscript.
MY supported the data processing and provided expertise on NIRCam data analysis.
JG contributed to determining the optimal pointing the mask for the second epoch based on the first epoch.
JL contributed to determining the optimal pointing the mask for the second epoch based on the first epoch.
AS provided expertise on converting the achieved contrast to planet masses and edited the manuscript.
MP, MM, LP, WB, AG, and KH contributed as a member of the NIRCam GTO team. All authors assisted in the preparation of the second epoch JWST proposal.

\end{contribution}

\facilities{JWST (NIRCam)}

\software{astropy \citep{astropy:2013, astropy:2018, astropy:2022}, spaceKLIP \citep{Kammerer22,Carter23}, BANYAN \citep{Gagne18}, 
MOCA \citep{Gagne26}.}

\appendix

\section{Planet Mass Limits}
\label{sec:appendix_mass_limits}

\begin{figure*}[htbp]
\centering
\includegraphics[width=0.49\textwidth]{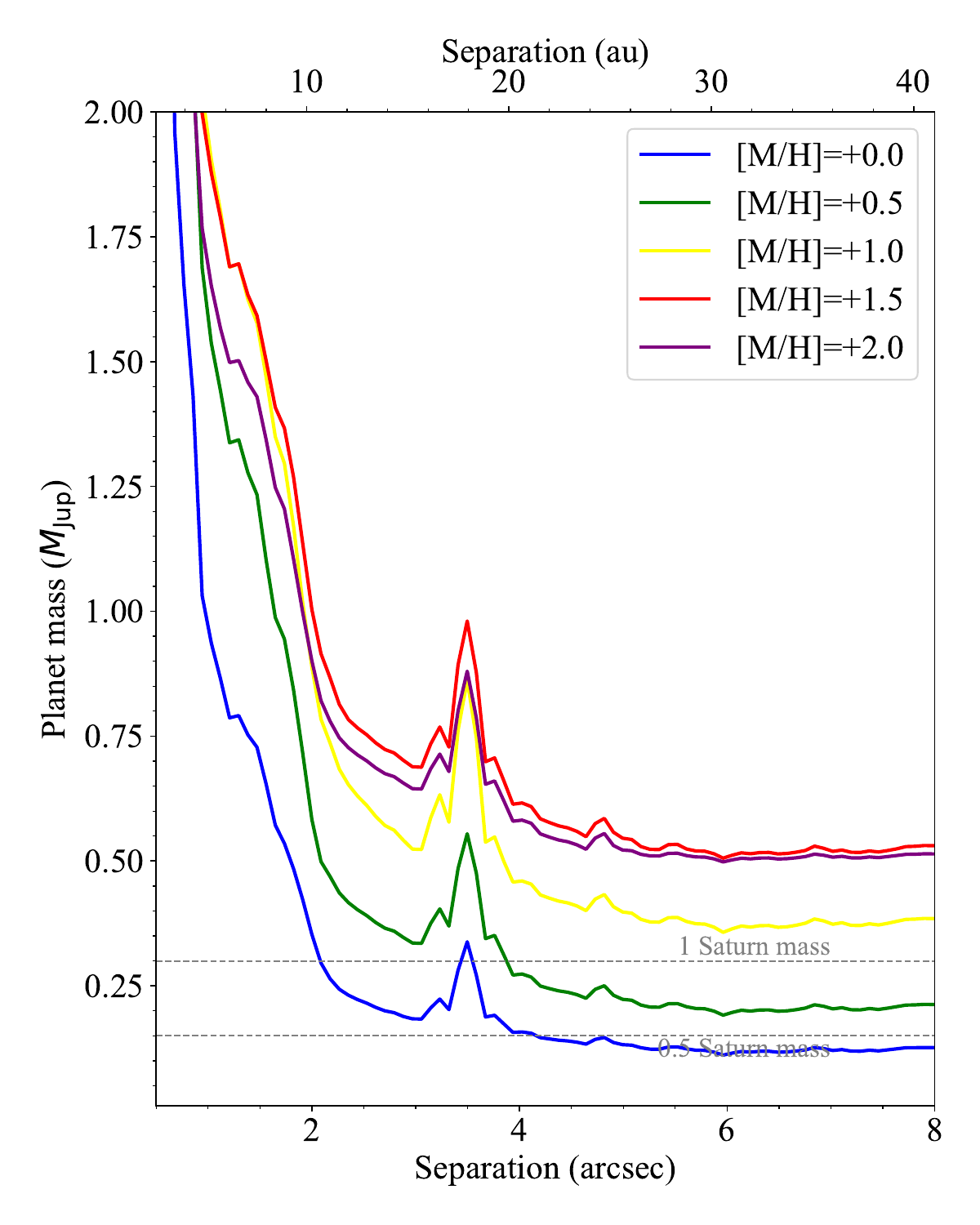}
\caption{A comparison of the planet mass sensitivity achieved by the NIRCam F444W Altair observations for varying atmospheric metallicities scenarios with the Sonora Flame Skimmer models (assuming strong vertical mixing and solar C/O ratio).}
\label{fig:flameskimmer_ap}
\end{figure*}

The conversion of the F444W contrast curve to a planetary mass sensitivity curve relies on the use of atmospheric and evolutionary models and requires assumptions to be made about the planet's atmospheric parameters, such as metallicity, vertical mixing strength, C/O ratio, and the presence/absence of clouds. A growing number of JWST observations and modeling efforts are now providing evidence that the 4--5~$\mu$m (F444W) flux of cold ($\lesssim$400~K), low-mass giant planets (such as the ones we are sensitive to in this work) can be dramatically lower due to a combination of strong vertical mixing and enhanced metallicity, which increase the strength of CO and CO$_2$ absorption, and the presence of water ice clouds, which act as an additional opacity source by redistributing flux to longer wavelengths \citep[e.g.,][]{2014ApJ...787...78M, 2023ApJ...950....8L, Crotts25, 2025ApJ...988L..18B, 2025ApJ...986L..26B, 2026AJ....171..225S, 2026AJ....172..115S}.

To account for this atmospheric model dependence, we additionally compute mass sensitivity curves for varying atmospheric metallicity (0~dex to +2.0~dex) and with strong vertical mixing (conservative case) using the clear Sonora Flame Skimmer models \citep{Mang2026}. Figure~\ref{fig:flameskimmer_ap} shows that the mass sensitivity degrades as the atmospheric metallicity increases, as expected. However, in all cases, we are still sensitive to $<1\;M_{\rm Jup}$ planets at separations $\gtrsim2\arcsec$. While the Sonora Flame Skimmer models do not include the effects of water clouds at cold temperatures, the highest metallicity case (+2.0~dex) can be used as a representative proxy since the impact on the F444W flux is similar. Note that Jupiter's atmosphere is enriched in metals at 3$\times$ (+0.5~dex) the solar value \citep{2004Icar..171..153W} and Saturn's atmosphere is enriched at the 10$\times$ (+1.0 dex) the solar value \citep{2009Icar..199..351F}, so it is unlikely that any giant planet around Altair is enriched at the +2.0~dex level.

\section{Background Star Astrometry and Photometry}
\label{sec:appendix_companions}

\begin{figure*}[htbp]
\centering
\includegraphics[width=0.49\textwidth]{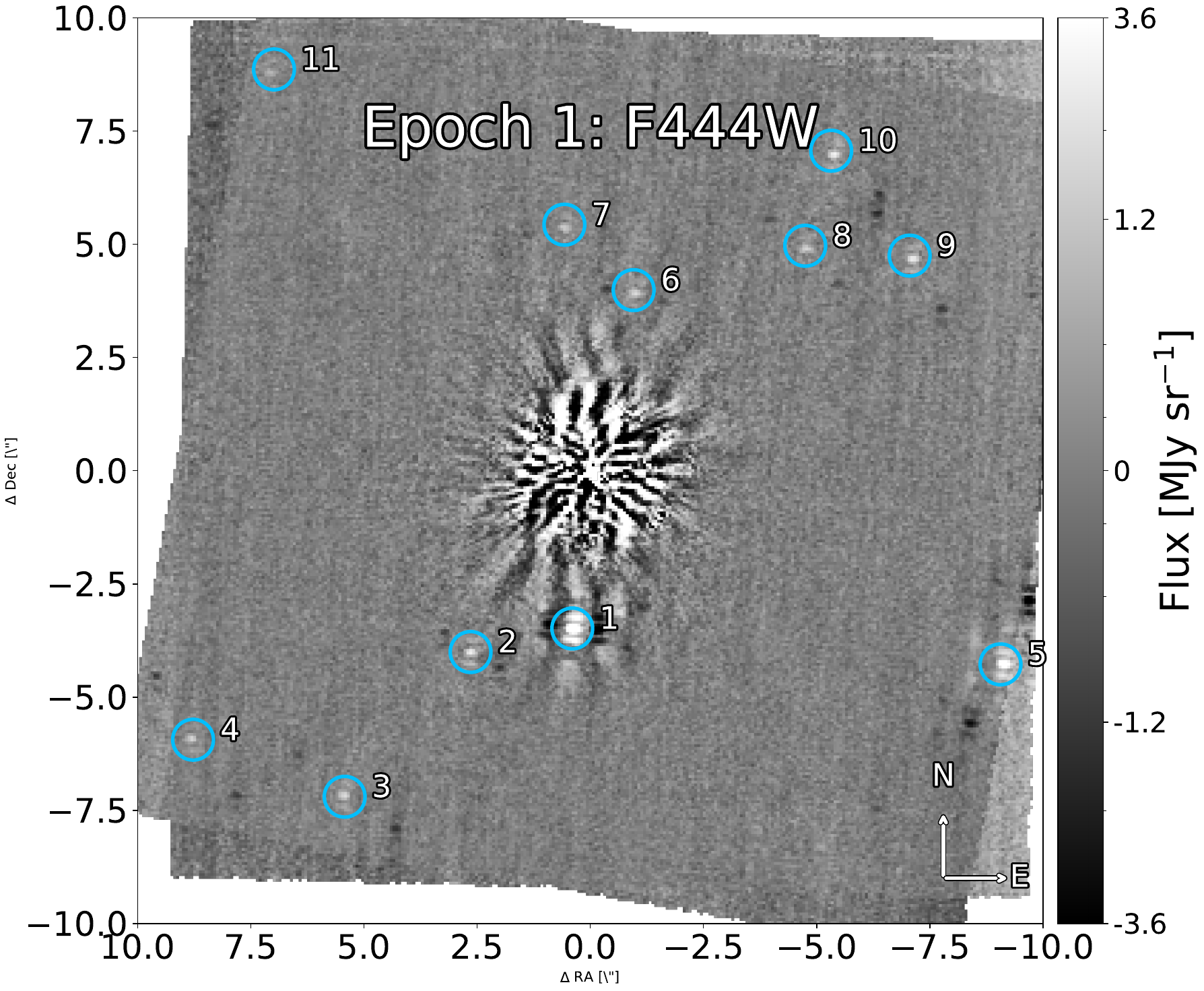}
\caption{KLIP-subtracted NIRCam coronagraphic images of the Altair field across both epochs and filters, with the eleven identified point sources circled and numbered. Object 1 is the bright background star discussed in the main text (Table~\ref{tab:bgstar}); objects 2--11 are presented in Table~\ref{tab:companions}. Not every object falls within the field of view of every epoch/filter combination.}
\label{fig:companion_finder}
\end{figure*}

In addition to the bright background star discussed in Section~\ref{subsec:bright_background_stars} (Table~\ref{tab:bgstar}), ten fainter point sources were identified in the field of view around Altair. This appendix presents their measured astrometry and photometry from spaceKLIP's \textit{extract\_companions} function (Table~\ref{tab:companions}), their F200W-F444W colors where simultaneous multi-filter coverage is available (Table~\ref{tab:companion_colors}), and their locations relative to Altair (Figure~\ref{fig:companion_finder}).

\begin{table*}[htbp]
\centering
\caption{Astrometry and photometry of additional candidate point sources in the Altair field. See Table~\ref{tab:bgstar} for the brightest background star (object 1).\label{tab:companions}}
\begin{tabular}{clccccll}
\hline\hline
ID & Date & Filter & $\Delta$RA ($''$) & $\Delta$Dec ($''$) & PA ($\degr$) & Sep. ($''$) & Contrast \\
\hline
\multirow{3}{*}{2}
    & 2024 Oct 11 & F200W & $+2.6727\pm0.0018$ & $-3.9933\pm0.0011$ & $146.21\pm0.02$ & $4.8052\pm0.0013$ & $9.69\pm0.20 \times 10^{-8}$ \\
    & 2024 Oct 11 & F444W & $+2.6445\pm0.0036$ & $-4.0039\pm0.0024$ & $146.56\pm0.04$ & $4.7984\pm0.0028$ & $1.37\pm0.03 \times 10^{-7}$ \\
    & 2025 May 26 & F444W & $+2.0394\pm0.0038$ & $-4.3350\pm0.0033$ & $154.81\pm0.04$ & $4.7907\pm0.0034$ & $1.06\pm0.03 \times 10^{-7}$ \\
\hline
\multirow{2}{*}{3}
    & 2024 Oct 11 & F444W & $+5.4262\pm0.0067$ & $-7.2008\pm0.0047$ & $143.00\pm0.04$ & $9.0164\pm0.0055$ & $1.20\pm0.04 \times 10^{-7}$ \\
    & 2025 May 26 & F444W & $+4.8433\pm0.0057$ & $-7.5137\pm0.0052$ & $147.19\pm0.04$ & $8.9395\pm0.0053$ & $8.07\pm0.35 \times 10^{-8}$ \\
\hline
\multirow{2}{*}{4}
    & 2024 Oct 11 & F444W & $+8.7754\pm0.0244$ & $-5.9408\pm0.0181$ & $124.10\pm0.11$ & $10.5972\pm0.0226$ & $6.69\pm1.08 \times 10^{-8}$ \\
    & 2025 May 26 & F444W & $+8.2051\pm0.0210$ & $-6.2746\pm0.0192$ & $127.41\pm0.11$ & $10.3293\pm0.0204$ & $8.41\pm1.21 \times 10^{-8}$ \\
\hline
\multirow{1}{*}{5}
    & 2024 Oct 11 & F444W & $-9.0636\pm0.0066$ & $-4.2770\pm0.0052$ & $244.74\pm0.03$ & $10.0221\pm0.0064$ & $1.89\pm0.11 \times 10^{-7}$ \\
\hline
\multirow{3}{*}{6}
    & 2024 Oct 11 & F444W & $-0.9580\pm0.0054$ & $+3.9877\pm0.0036$ & $346.49\pm0.07$ & $4.1011\pm0.0037$ & $1.13\pm0.04 \times 10^{-7}$ \\
    & 2025 May 26 & F444W & $-1.6311\pm0.0076$ & $+3.6306\pm0.0060$ & $335.81\pm0.11$ & $3.9802\pm0.0063$ & $1.00\pm0.05 \times 10^{-7}$ \\
    & 2025 May 26 & F200W & $-1.6513\pm0.0021$ & $+3.6172\pm0.0017$ & $335.46\pm0.03$ & $3.9763\pm0.0018$ & $6.43\pm0.19 \times 10^{-8}$ \\
\hline
\multirow{3}{*}{7}
    & 2024 Oct 11 & F444W & $+0.5721\pm0.0059$ & $+5.4323\pm0.0036$ & $6.01\pm0.06$ & $5.4623\pm0.0037$ & $1.03\pm0.03 \times 10^{-7}$ \\
    & 2025 May 26 & F444W & $-0.0504\pm0.0087$ & $+5.0926\pm0.0069$ & $359.43\pm0.10$ & $5.0928\pm0.0069$ & $4.67\pm0.29 \times 10^{-8}$ \\
    & 2025 May 26 & F200W & $-0.0951\pm0.0058$ & $+5.0586\pm0.0050$ & $358.92\pm0.07$ & $5.0595\pm0.0050$ & $1.90\pm0.18 \times 10^{-8}$ \\
\hline
\multirow{3}{*}{8}
    & 2024 Oct 11 & F444W & $-4.7495\pm0.0058$ & $+4.9669\pm0.0044$ & $316.28\pm0.04$ & $6.8722\pm0.0051$ & $8.73\pm0.28 \times 10^{-8}$ \\
    & 2025 May 26 & F444W & $-5.3676\pm0.0058$ & $+4.6723\pm0.0047$ & $311.04\pm0.04$ & $7.1162\pm0.0053$ & $8.09\pm0.27 \times 10^{-8}$ \\
    & 2025 May 26 & F200W & $-5.4187\pm0.0035$ & $+4.6191\pm0.0030$ & $310.45\pm0.03$ & $7.1203\pm0.0033$ & $3.08\pm0.14 \times 10^{-8}$ \\
\hline
\multirow{3}{*}{9}
    & 2024 Oct 11 & F444W & $-7.0564\pm0.0052$ & $+4.7487\pm0.0035$ & $303.94\pm0.03$ & $8.5055\pm0.0047$ & $8.95\pm0.35 \times 10^{-8}$ \\
    & 2025 May 26 & F444W & $-7.7285\pm0.0040$ & $+4.4462\pm0.0032$ & $299.91\pm0.02$ & $8.9162\pm0.0038$ & $1.33\pm0.03 \times 10^{-7}$ \\
    & 2025 May 26 & F200W & $-7.7626\pm0.0014$ & $+4.4104\pm0.0014$ & $299.60\pm0.01$ & $8.9280\pm0.0014$ & $1.19\pm0.03 \times 10^{-7}$ \\
\hline
\multirow{3}{*}{10}
    & 2024 Oct 11 & F444W & $-5.3220\pm0.0049$ & $+7.0661\pm0.0034$ & $323.01\pm0.03$ & $8.8461\pm0.0040$ & $1.08\pm0.04 \times 10^{-7}$ \\
    & 2025 May 26 & F444W & $-6.0054\pm0.0051$ & $+6.7539\pm0.0042$ & $318.36\pm0.03$ & $9.0377\pm0.0046$ & $1.00\pm0.04 \times 10^{-7}$ \\
    & 2025 May 26 & F200W & $-6.0328\pm0.0058$ & $+6.7122\pm0.0044$ & $318.05\pm0.03$ & $9.0249\pm0.0051$ & $7.28\pm0.42 \times 10^{-8}$ \\
\hline
\multirow{2}{*}{11}
    & 2024 Oct 11 & F444W & $+6.9878\pm0.0346$ & $+8.8614\pm0.0381$ & $38.26\pm0.18$ & $11.2851\pm0.0368$ & $4.34\pm1.08 \times 10^{-8}$ \\
    & 2025 May 26 & F444W & $+6.4505\pm0.0471$ & $+8.5213\pm0.0366$ & $37.13\pm0.23$ & $10.6875\pm0.0407$ & $1.10\pm0.20 \times 10^{-7}$ \\
\hline
\end{tabular}
\begin{minipage}{\textwidth}
\smallskip
\textit{Note.} Astrometry and contrast measured from JWST/NIRCam coronagraphic imaging using \texttt{spaceKLIP}. Positive $\Delta$RA is East, positive $\Delta$Dec is North. Objects 2--4 and 6--10 show consistent apparent motion relative to Altair between epochs ($\sim$0.7$''$ over $\sim$7.5 months, in a common direction), confirming their identification as background stars rather than physically associated companions. Objects 3 and 4 fall outside the field of view of the F200W Epoch 1 observation. Object 5 is detected in only one epoch/filter combination and its background/companion status remains unconstrained. Object 11's motion is broadly consistent with the background-star group, though its low SNR ($\sim$2--5) makes this less certain.
\end{minipage}
\end{table*}

\begin{table}[htbp]
\centering
\caption{Background object F200W$-$F444W colors and apparent motion relative to Altair between epochs \label{tab:companion_colors}}
\begin{tabular}{lcc}
\hline\hline
ID & F200W$-$F444W & Motion ($''$, PA) \\
\hline
1  & $+0.27\pm0.07$ & $0.70$, $242.7\degr$ \\
2  & $+0.39\pm0.08$ & $0.69$, $241.3\degr$ \\
3  & \ldots         & $0.66$, $241.8\degr$ \\
4  & \ldots         & $0.66$, $239.7\degr$ \\
6  & $+0.50\pm0.10$ & $0.76$, $242.1\degr$ \\
7  & $+0.99\pm0.14$ & $0.71$, $241.4\degr$ \\
8  & $+1.06\pm0.10$ & $0.68$, $244.5\degr$ \\
9  & $+0.14\pm0.08$ & $0.74$, $245.8\degr$ \\
10 & $+0.37\pm0.11$ & $0.75$, $245.5\degr$ \\
11 & \ldots         & $0.64$, $237.7\degr$ \\
\hline
\end{tabular}
\begin{minipage}{\columnwidth}
\smallskip
\textit{Note.} Colors are only computed when there is simultaneous F200W and F444W measurements. Objects 3, 4, and 11 do not have simultaneous F200W/F444W coverage due to the limited FOV of F200W. Object 5 (only detected in Epoch 1, F444W) is removed from this table, as no motion or color measurement could be computed. ``Motion" is the magnitude and position angle (E of N) of each object's apparent motion relative to Altair between the two epochs ($\sim$7.5 months apart), derived from Table~\ref{tab:companions}. The spread in color ($+0.14$ to $+1.06$ mag) is consistent with a population of background stars of varying spectral type. The consistent motion across all sources ($0.64$-$0.76''$, $238$--$246\degr$) confirms their status as stationary background objects rather than physically associated companions.
\end{minipage}
\end{table}


\bibliography{meshkat.bib}{}
\bibliographystyle{apj}

\end{document}